# Characterization of spatially inhomogeneous chirp in ultrashort multielectron beams via femtosecond hole burning

Yuichi Tachibana[1,2*] and Yuya Morimoto[1,3*]

[1]*RIKEN Center for Advanced Photonics (RAP), RIKEN, 2-1 Hirosawa, Wako, Saitama 351-0198, Japan*

[2]*Department of Physics, Tokyo University of Science, 1-3 Kagurazaka, Shinjuku, Tokyo 162-8601, Japan*

[3]*RIKEN Pioneering Research Institute (PRI), RIKEN, 2-1 Hirosawa, Wako, Saitama 351-0198, Japan*

*yuichi.tachibana@rs.tus.ac.jp;

*yuya.morimoto@riken.jp

**Direct observation of the temporal–energy structure of pulsed electron beams is crucial for beam-driven light generation and time-resolved microscopic imaging. In this study, we characterized the spatially non-uniform time–energy structure in multi-electron ultrashort pulses exhibiting significant space-charge effects. By combining spectral hole burning originating from the photon-induced near-field electron microscopy (PINEM) effect with angle-resolved energy analysis, we observed time–energy correlations at each beam angle. Applying this method to 37-keV pulses containing up to 17 electrons per pulse, we determined energy spread and chirp rate, which vary by up to 100% and 40%, respectively, across the entire beam. These findings highlight the importance of characterizing multi-electron pulses with spatial or angular resolution and suggest the utility of spatially modulated light for the shaping of high-flux pulsed electron beams.**

## 1. Introduction

Characterizing the temporal structure of pulsed electron beams is crucial for various electron beam applications [1]. In X-ray free-electron lasers, attosecond X-ray pulses have been generated by controlling electron phase-space structures [2]. Similarly, in ultrafast transmission electron microscopy, precise manipulation of time and energy using light waves has enabled electromagnetic imaging with attosecond temporal resolution [3–7].

For pulsed electron beams at relativistic speeds, radio-frequency (RF) or terahertz (THz) deflectors [8–11], electro-optic sampling [12], and coherent radiation [13–15] have been employed to observe their temporal structures. Through these methods, the temporal information of the electrons is converted into spatial profiles on an electron detector, emission light spectra, or the polarization state of probe light. Similarly, pulsed electrons at sub-relativistic speeds are often characterized by inducing ultrafast changes in the beam's spatial profile on a detector [16]. Techniques using ultrashort laser pulses, such as beam scattering via the ponderomotive potential [17–21], plasma lensing [22], and streaking using terahertz light [23–28] have been developed. Furthermore, the temporal resolution of streaking measurements was recently enhanced to attoseconds by utilizing electromagnetic fields at higher (i.e., optical) frequencies [29–35].

Extending time-resolved measurements to capture time–energy correlations yields deeper insights into dispersion and space-charge effects. In ultrafast electron microscopy, the coherent energy modulation of electron beams via light–electron coupling, known as the photon-induced near-field electron microscopy (PINEM) effect [36], has been widely exploited. For monochromatic electron beams, energy shifts corresponding to integer multiples of the photon energy were observed [36,37]. By leveraging energy changes from light-electron interactions instead of the spatial beam modulation, this technique successfully characterized both the pulse duration and chirp [38]. The reduction in the intensity of the intrinsic energy component (i.e., zero-loss peak intensity) caused by the PINEM effect was also used to measure the electron-beam pulse duration [39–41]. Conversely, for electron beams with broad energy spreads or with energy resolution insufficient to resolve the photon-induced variations, spectral hole burning, where the pre-existing spectral components decrease due to the interaction with light, has been employed to measure the correlation between temporal and energy structures [38,42].

Prior studies have typically characterized the pulse duration and energy structure as averages over space or beam angle. However, an electron beam inherently possesses a finite beam size and angular distribution, resulting in a non-uniform intensity distribution. When the average number of electrons per pulse is small enough that Coulomb repulsion between electrons can be neglected, the spatial inhomogeneity of pulse duration and energy spread can also be negligible. However, when the number of electrons per pulse is large and the space-charge effect (i.e., Boersch effect) is significant, broadening in both energy and time occurs in correspondence with the electron flux [43–47]. In particular, for spatially localized sources such as nano-tips, strong Coulomb repulsion acts during electron emission [48–52], which can lead to spatially non-uniform time and energy structures. For applications such as monochromatization [53,54], chromatic aberration correction [55], temporal compression [11,23,24], or three-dimensional pulse shaping [56] using light waves, it is highly beneficial to investigate the relationship between time and energy across the whole beam profile.

In this study, we report the characterization of the spatially inhomogeneous time–energy structure of an electron beam obtained from a tip emitter by applying angle-resolved energy spectroscopy to the PINEM-based spectral hole burning. The ultrashort electron pulses containing multiple electrons exhibit spatially non-uniform energy and chirp distributions which are highly correlated with the spatial profile of the beam intensity.

## 2. Experimental methods

Figure 1a shows a schematic of the experimental setup. The second harmonic of the output of a femtosecond laser (1033 nm, 190 fs, 200 kHz) was used for the generation of pulsed electrons from a tungsten needle-type emitter (Denka) via two-photon emission. The apex of the emitter has a diameter of several micrometers, and no direct-current (DC) field emission was observed. The emitted electrons were accelerated to a kinetic energy of 37 keV by an electrostatic field. The electron beam was collimated and then spatially focused by a set of magnetic lenses. Therefore, the spatial profile of the beam was converted into an angular distribution. No crossover was formed between the lenses. A slit with dimensions of 0.125 mm × 1 mm was installed between the lenses to enhance the resolution of the spectral measurement. The orientation of the slit was adjusted such that the beam became vertically elongated (i.e., along the $y$-direction) both before and after the focal point. The energy spectrum of the electron beam was observed using a custom-built spectrometer with a sector magnet and magnetic multipoles [57]. An angle-resolved energy spectrum was obtained as an image, as shown in Figs. 1b and 1c, with an energy resolution of 3.2 eV in full width at half maximum (FWHM). The angular resolution is about 0.1 mrad, which corresponds to a spatial resolution of ~0.02 mm at the position of the slit. Due to the misalignment of the anode hole (0.8 mm in diameter), the spatial profile of the beam was asymmetric, as shown by the profile on the beam camera (inset of Fig. 1a). This asymmetry also manifests in the angular distribution of the energy spectrum (i.e., along the vertical direction in Figs. 1b and 1c).

The temporal–energy correlations of the pulsed electron beam were observed through the energy modulation induced by the ultrashort laser field (190 fs, 1033 nm). A 20-nm-thick $Si_3N_4$ membrane (Norcada) was employed to facilitate coupling between the beamed electrons and the laser field [29,58–60]. The membrane was placed near the focal point of the electron beam with a spot size of 16 μm ($x$) × 58 μm ($y$). At the beam flux used in this study, there were no significant changes in the spot size and, accordingly, the transverse emittance. The laser beam was loosely focused (245 μm ($x$) $\times$ 288 μm ($y$), 1/e$^2$ full width). By tilting the membrane by 20 degrees relative to the normal incidence of the electron beam, velocity matching between the electrons and the laser light along the $x$-direction was satisfied, enabling temporal measurement independent of the spatial size of the beam and free from field-driven deflection [29,61].

## 3. Results and discussion

To examine the strength of the electron–light coupling, we first compare electron energy spectra observed with a low-flux electron beam (1 electron/pulse), where the space-charge effect is negligibly small. Figure 2a shows the spectra with (red) and without (black) the laser field (0.5 V/nm peak amplitude). Here, the signal was integrated over the beam angle (from –2.3 mrad to 2.3 mrad). A significant broadening of the spectral widths (red curve) to 8.2 eV (FWHM), corresponding to an energy gain/loss of up to ~3 photons, was observed only when the electron and laser pulses overlapped both spatially and temporally. Owing to the limited energy resolution of our apparatus (3.2 eV), a continuous profile rather than discrete peaks was obtained.

In contrast, when there are multiple electrons in a pulse, the intrinsic width of the energy distribution is broadened due to the space-charge effect [43–52]. Figure 2b shows the relationship between the average number of electrons in a pulse (without the slit) and the observed energy width in FWHM. We note that the flux here was measured using a Faraday cup in the absence of the membrane and the slit. Up to 2 electrons/pulse, the energy width is comparable to or smaller than the resolution (3.2 eV). Exceeding 4 electrons per pulse results in a steep increase in energy width. Figure 2c presents the spectra with a beam with an average flux of 10 electrons/pulse and an energy width of 10 eV (FWHM). The electron-laser delay time is fixed. From top to bottom, the laser peak electric field amplitudes are 0 V/nm, 0.2 V/nm, 0.3 V/nm, and 0.5 V/nm. No clear overall spectral broadening was observed even at the maximum field strength of 0.5 V/nm. Instead, a distinct dip was observed near an energy shift of 0 eV. This spectral hole burning effect has previously been observed in electron microscopes [38–42] and occurs when only the specific spectral components of the incident electron beam interact with the laser, causing a depletion of intensity at that spectral component. Furthermore, spatio-temporal mapping of the electric field has also been achieved by leveraging the hole burning [62].

The depth of the dip observed in this experiment increased with an increase in laser field amplitude. The probability of absorbing or emitting $n$ photons is represented by $J_n^2(2|g|)$ [63–65], where $J_n(x)$ is the Bessel function of the first kind and $2|g|$ is a parameter representing the coupling strength between the light and the electron, corresponding to the maximum classically allowed energy modulation normalized by the photon energy. At an electric field strength of 0.5 V/nm ($2|g|{\sim}3$ as seen above), $J_0^2(3) \cong 0.07$ which leads to an expected dip depth of nearly 100%. However, due to the limited energy resolution, the observed signal depletion was limited to approximately 30%. The energy width of the dip (~2 eV) sets the upper limit of the instantaneous energy spread of the electrons.

Figure 2d shows the temporal evolution of specific energy components as a function of the electron-laser delay time. The signal intensity at an energy 2 eV above the center energy (top blue curve) exhibits a dip at positive delay times, whereas that at an energy −2 eV (bottom yellow curve) shows a dip at negative delay times. This indicates a linear chirp, where higher-energy electrons arrive at the membrane at earlier times. Here, the temporal width of the dip is ~400 fs (FWHM) across all three curves in Fig. 2d, showing a slower response than the laser pulse duration (190 fs). This temporal broadening originates from the finite energy resolution (3.2 eV). For a chirp rate of 5 meV/fs (see below), a response time of ~600 fs is expected. Thus, achieving higher temporal resolution requires a higher energy resolution.

The data presented above were obtained by integrating the signals over the convergence angle of the beam. In our experiment, angle-resolved energy spectra, as shown in Fig. 1b-c, were acquired. We note that signals are averaged along the *y*-direction (i.e., perpendicular to the slit). Representative results are shown in Fig. 3a, corresponding to beam fluxes of 4, 10, and 17 electrons/pulse (from top to bottom) with energy spreads (FWHM) of 5 eV, 10 eV, and 15 eV, respectively. The topmost image in each stacked plot represents the signal integrated over an angular range of $0 \pm 0.25$ mrad. In all images, diagonal traces corresponding to the spectral dips extend from the lower-left to the upper-right region, showing a linear chirp.

From the time-resolved spectra obtained at each beam angle (Fig. 3a), the time–energy correlation of the electron beam can be extracted at each beam angle. Figure 3b shows the mean energy (i.e., the dip position in the spectrum) at each delay time and beam angle. Notably, in all cases, the energy distribution and associated temporal width are broader in the negative angular region (lower side) compared to the positive angular region (upper side). At 17 electrons/pulse (lower panel), both the energy width and the temporal width at negative angles are approximately twice as large as those at positive angles. This behavior correlates with the non-uniformity of the beam intensity, as shown in Fig. 1a, where the intensity is stronger in the negative angular region. Furthermore, this non-uniformity in the time–energy structure was more pronounced at higher flux, which is attributed to the non-linear increase in energy spread with respect to the number of electrons per pulse (Fig. 2b). Such spatial non-uniformity of the energy and temporal structures within the beam is typically averaged out in conventional measurements that lack angle or position resolution.

The dispersion of the multi-electron pulse is investigated in more detail. Figure 4a shows the variation of the electrons' instantaneous mean energy within the angular range of $0 \pm 0.25$ mrad. For all three electron fluxes shown, linear chirp is dominant. In addition, the chirp rate (slope) at 4 electrons/pulse (black circles) is noticeably smaller than those at the higher fluxes. We then investigate the non-uniformity of the chirp across the beam profile. The upper panel of Fig. 4b displays the chirp rate at each beam angle observed with 4 electrons/pulse. It shows that the chirp rate varies within the beam from 2.8 meV/fs to 4.1 meV/fs, representing a variation of more than 40%. When the chirp rate (upper panel) is compared to the energy spread (lower panel), we find a large correlation between them. Specifically, a larger energy spread corresponds to a higher chirp rate. For comparison, Figs. 4c and 4d present the corresponding results at 10 and 17 electrons/pulse, respectively. Although the variation of the energy width with angle is larger, the non-uniformity of the chirp is smaller with the higher flux, varying by 15% (from 4.9 meV/fs to 5.7 meV/fs) for 10 electrons/pulse and 13% (from 5.2 meV/fs to 6 meV/fs) for 17 electrons/pulse. However, clear dependences on the beam angle and energy width are still observed.

Black circles in Fig. 5 show the correlation between the energy spread and the chirp rate observed across various fluxes and beam angles, including the data shown in Figs. 4b–4d. Remarkably, the data points measured at different electron fluxes and angles fall onto a single unified curve. At relatively small energy spreads (5–10 eV), the chirp rate increases sharply, whereas for larger energy spreads (>10 eV), the variation becomes less pronounced. The nearly constant chirp rate at large energy spreads shows that the dispersion occurring during propagation in vacuum predominantly determines the time–energy structure within the pulse, regardless of the initial pulse duration upon emission. Due to the non-linear dependence of chirp rate on the energy spread, the high spatial inhomogeneity of the chirp rate was observed at the intermediate energy spreads in Fig. 4(b). On the other hand, a higher uniformity of the chirp rate was achieved with a multi-electron pulse containing a larger number of electrons exhibiting a broader energy spread.

Assuming a Gaussian temporal-spectral structure, the chirp rate $\chi$ is given approximately by the following equation [49],

$$\chi(\sigma_E) = \frac{\sigma_E^2 s_E}{\sigma_E^2 s_E^2 + \sigma_t^2} \quad (1),$$

where $\sigma_E$ represents the electron energy width, $s_E$ is the energy-dependent difference in arrival time at the sample (here, a membrane), and $\sigma_t$ is the initial temporal width of the electrons. In this model, a two-step process is assumed: the energy width expands immediately after emission from the emitter due to strong space-charge effects, followed by pulse chirping occurring during flight. The solid curve in Fig. 5 shows the fitting result calculated using Eq. (1), which accurately reproduces the experimental data. For simplicity, experimental spectra were assumed to be Gaussians. From the fitting analysis, we obtained $s_E = 164 \pm 1$ fs/eV and $\sigma_t = 273 \pm 4$ fs. Since the chirp accumulated during the drift (a path length of ~700 mm) after acceleration to 37 keV is merely 79 fs/eV, the chirp induced during acceleration inside the electron gun makes a considerable contribution.

The dispersion of the ultrashort multi-electron pulses observed in this study is dominated by linear effects, indicating that these pulses can be monochromatized using terahertz (THz) waves [54] with appropriate frequency and amplitude. However, to achieve high monochromaticity, the THz wave must possess a spatial–amplitude or spatial–frequency correlation tailored to the varying chirp rates across different angular components (or spatial positions) of the beam. Furthermore, the maximum chirp rate observed in this experiment is 6 meV/fs. When applying attosecond temporal compression using optical waves, the energy spread originating from the chirp within an optical cycle (e.g., 3.4 fs for a wavelength of 1030 nm) is several tens of meV. In addition, the instantaneous energy spread at each moment (Fig. 2c) is below the energy resolution of our analyzer (3.2 eV). Therefore, sub-optical-cycle temporal modulation can be imparted to multi-electron pulses with an energy modulation amplitude (e.g., a few eV) significantly smaller than the overall energy spread of the beam (e.g., 15 eV in FWHM) [57].

## 4. Conclusion

We revealed the spatially non-uniform time–energy correlation within an electron beam by integrating angle-resolved energy spectroscopy into the temporal diagnosis based on PINEM-induced spectral hole burning. In multi-electron pulses where the space-charge effect is pronounced, large non-uniformities in both the energy spread and the chirp rate were observed, and they are strongly correlated. Notably, the chirp rate varied by up to 40% within the beam. Therefore, for ultrashort electron beams containing multiple electrons, temporal measurements that integrate over beam angle or spatial position can be insufficient, highlighting the need for angle- or position-resolved characterization. The time resolution of ultrafast electron microscopy and diffraction using multi-electron pulses may depend on the beam position and angle. Furthermore, to precisely manipulate such pulsed electron beams, it is necessary to apply modulation tailored in space and time.

As a future prospect, extending the method presented here by scanning the slit along the *y*-direction (see Fig. 1a) will enable the reconstruction of a full three-dimensional time–energy correlation of a pulsed electron beam. Moreover, combining a shorter laser pulse with a higher-resolution energy analyzer would allow for the detailed, energy-resolved observation of spatio-temporal structures in even shorter electron pulses, such as those temporally compressed by microwaves [18,66] or terahertz waves [11,23,24].

**Acknowledgements**

This research was supported by MEXT/JSPS KAKENHI JP21K21344, JP25K22230, JP 25K01734, JST FOREST JPMJFR2228, and Gordon and Betty Moore Foundation. We would like to thank R-COMS_RAP (Advanced Manufacturing Support) for their support in manufacturing components.

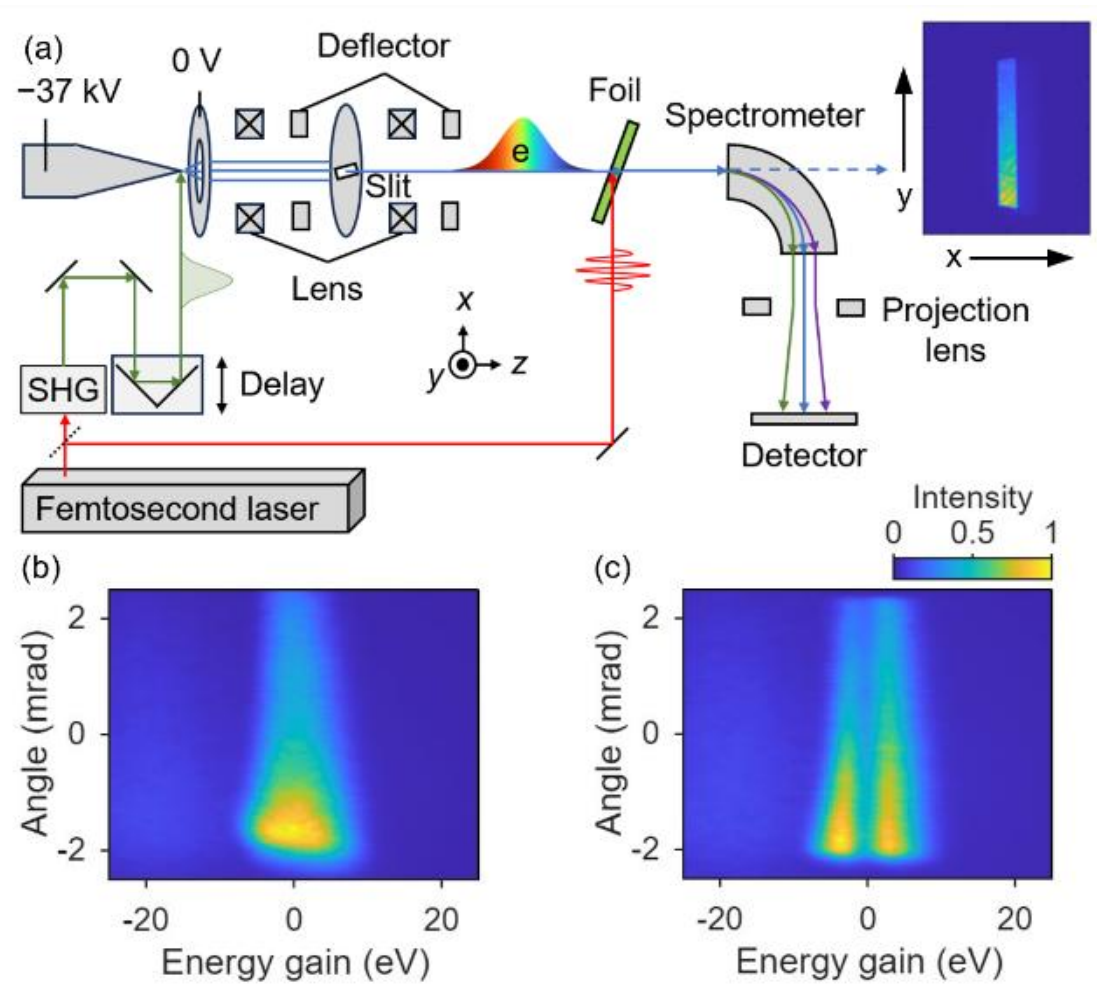

**Fig. 1.** Experimental overview. (a) Schematic of apparatus. An ultrashort multi-electron pulse was produced via a two-photon photoelectric effect at the apex of a tungsten needle emitter. The emitted electrons were accelerated to 37 keV by an electrostatic field. The electron beam was spatially focused near a 20-nm-thick silicon nitride membrane placed 700 mm downstream. Spatial-temporal overlap of the electron and laser pulses at the membrane resulted in energy modulation. The membrane was tilted about 20 degrees to satisfy the electron-light velocity matching condition along the *x*-direction. An angle-resolved electron-energy spectrum was observed using a home-made spectrometer. A 125-μm slit was employed to improve energy resolution. The top-right inset shows the two-dimensional spatial profile of the beam recorded by a CMOS camera. (b), (c) Observed angle-resolved energy spectra of the multi-electron beam (b) without and (c) with the temporal overlap with the laser pulse. Energy gain is expressed as the shift from 37 keV. Spatial energy-width inhomogeneity in both (b) and (c) and the spectral hole burning effect in (c) are clearly visible.

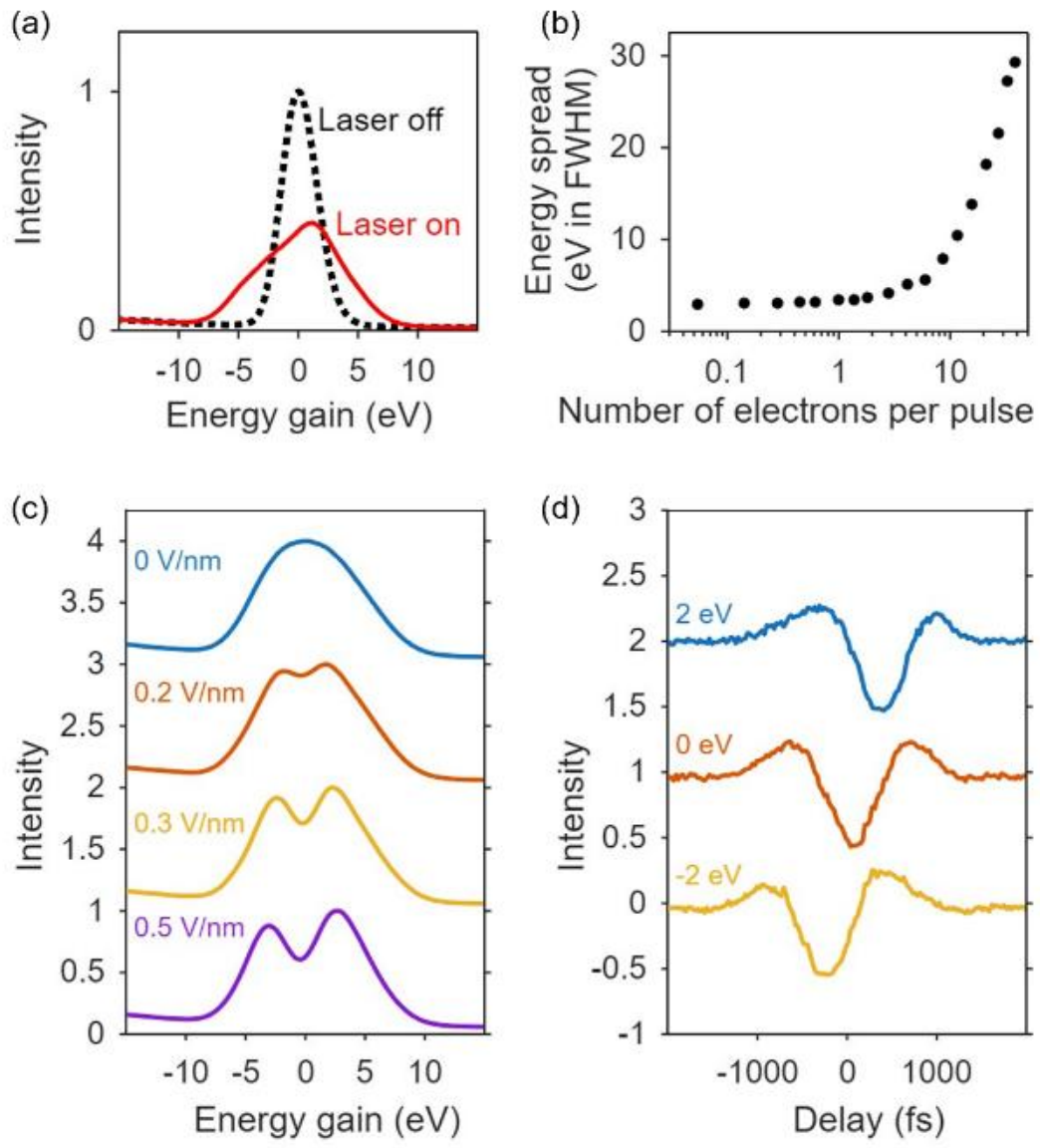

**Fig. 2.** Analysis of spatially integrated signals. (a) Electron-energy spectra for an average flux of one electron/pulse, where space-charge effects are negligibly small. (b) Energy broadening induced by the space-charge effect. A pronounced increase in energy spread was observed when the electron flux exceeded four electrons/pulse (without the slit). The reported energy spread includes the analyzer resolution. (c), (d) Results obtained at a flux of 10 electrons/pulse with an initial energy spread of 10 eV (FWHM). (c) Evolution of the energy spectrum as a function of peak laser electric field strength at a fixed delay time. (d) Delay dependence of the signal intensity at specific final electron energies.

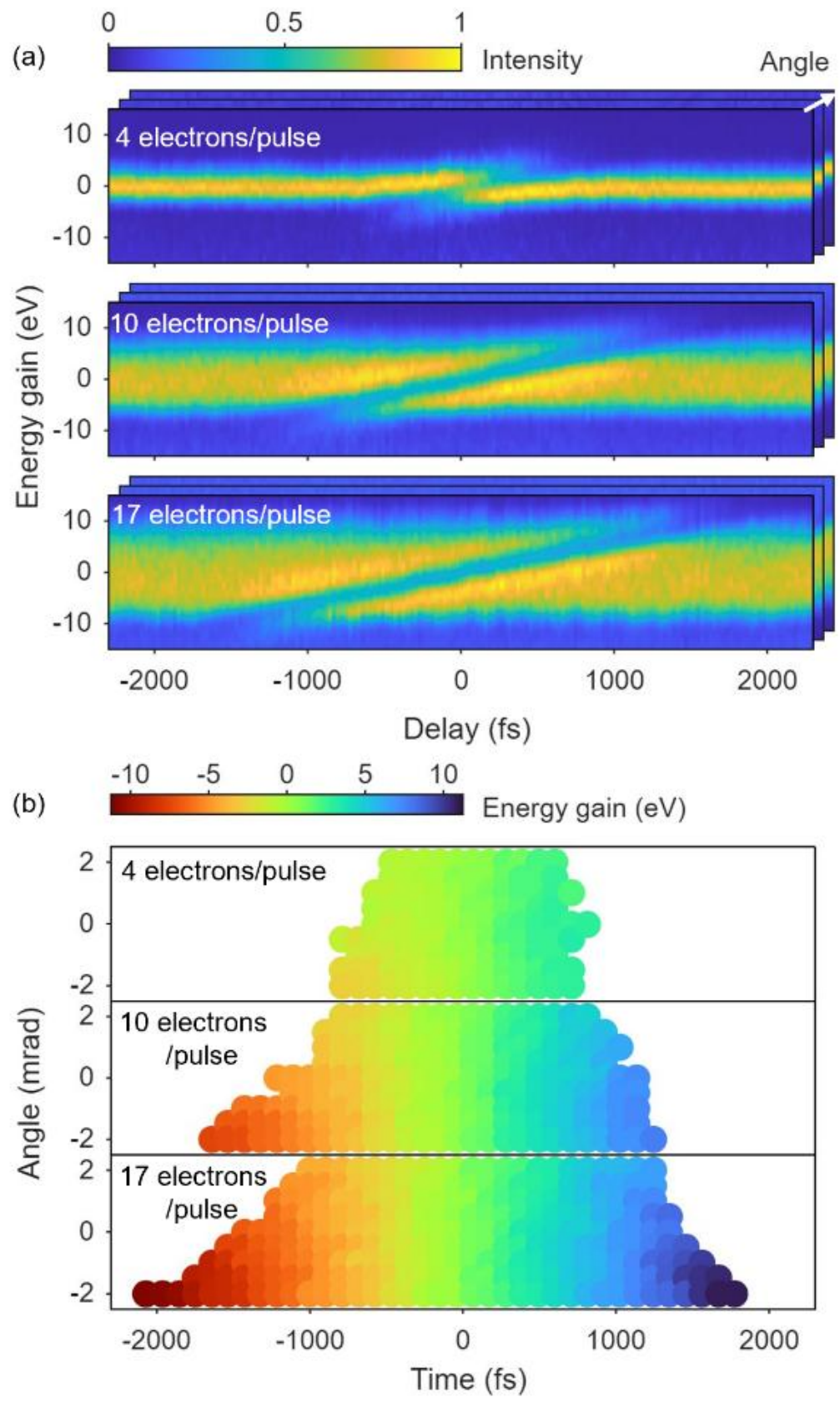

**Fig. 3.** Results of angle-resolved spectrum measurements. (a) Delay-dependent electron-energy spectra at each beam angle and flux. The top, middle, and bottom panels correspond to results for fluxes of 4, 10, and 17 electrons/pulse, respectively. (b) Angle-resolved time–energy structure of pulsed multi-electron beams retrieved via spectral hole burning.

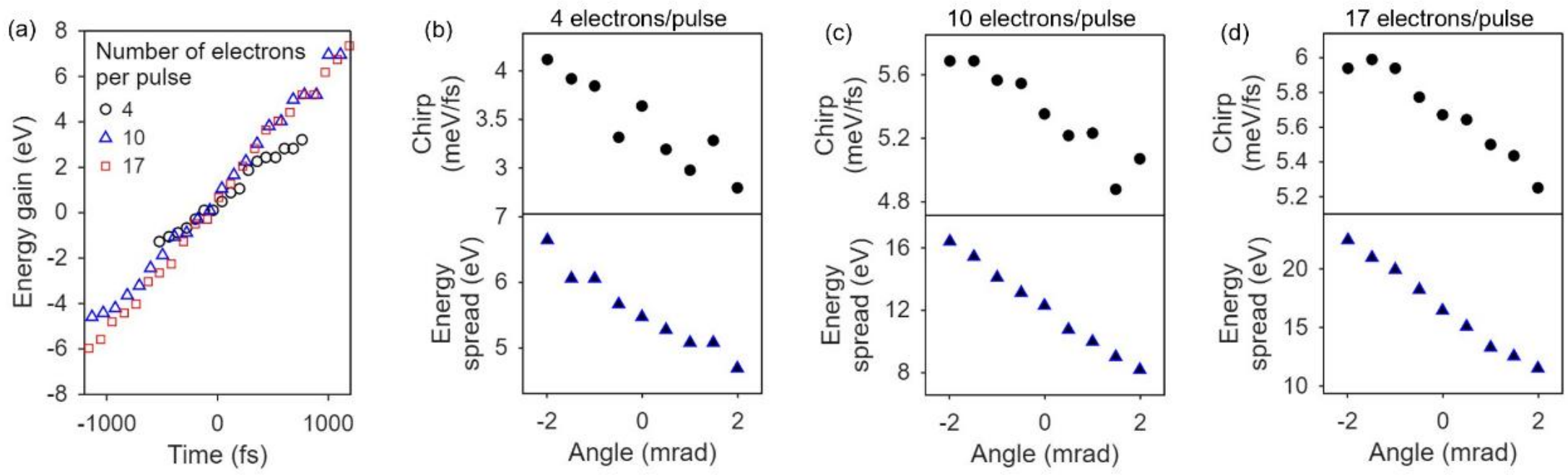

**Fig. 4.** Dispersion of multi-electron pulses. (a) Time–energy correlation within the angular range of 0 ±0.25 mrad. (b)–(d) Upper panels: chirp rates at each beam angle. Lower panels: energy spreads (in FWHM). The reported energy spread includes the analyzer resolution. Panels (b), (c), and (d) correspond to results obtained for average fluxes of 4, 10, and 17 electrons/pulse, respectively.

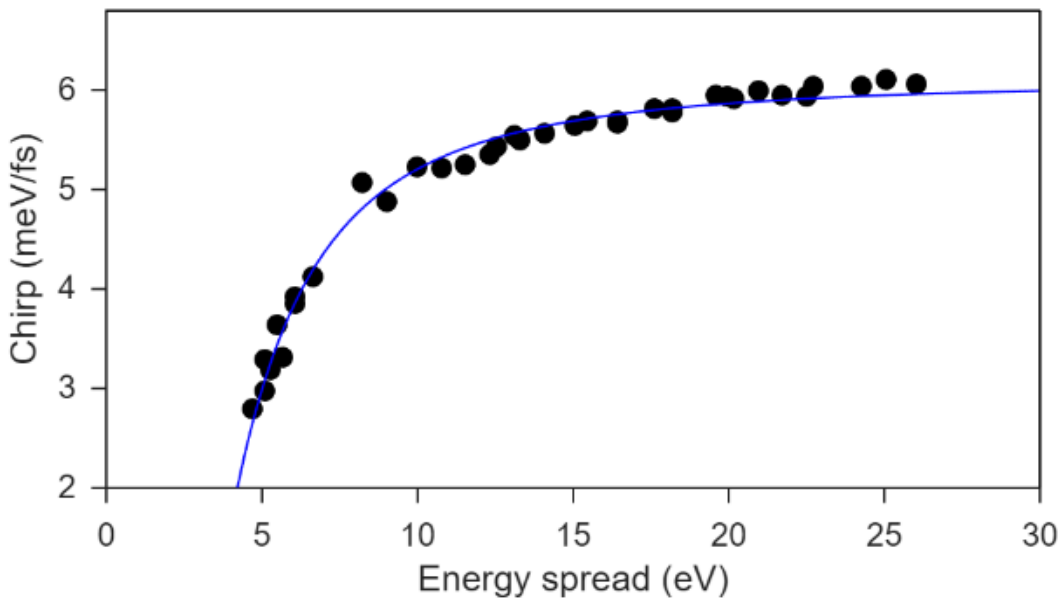

**Fig. 5.** Correlation between energy spread (in FWHM) and chirp rate. Data in Figs. 4b–d are shown alongside other observations. The reported energy spread includes the analyzer resolution (3.2 eV). Black circles: observed values. Blue line: fitted curve using Eq. (1).